\documentclass{elsart-1}

 \usepackage{graphicx}
 \usepackage{epsfig}

\usepackage{amssymb}

\usepackage[english,francais]{babel}

 \usepackage{epsfig}
 \usepackage{color,graphics}
 \usepackage{ latexsym,amsmath,amsfonts,amssymb}
 \usepackage{euscript,amsbsy}
 \usepackage{graphicx}
 \usepackage{caption}

\usepackage{latexsym}
\usepackage{amscd}
\usepackage{color,graphics}
\usepackage{graphicx}

\newtheorem{e-proposition}[theorem]{Proposition}

\newtheorem{e-definition}[theorem]{Definition\rm}

\renewcommand{\theequation}{\arabic{equation}}
\def\og{\leavevmode\raise.3ex\hbox{$\scriptscriptstyle\langle\!\langle$~}}
\def\fg{\leavevmode\raise.3ex\hbox{~$\!\scriptscriptstyle\,\rangle\!\rangle$}}

\newcommand{\bb}[1]{{\mathbb{#1}}}

\newcommand{\bea}{\begin{eqnarray}}
\newcommand{\eea}{\end{eqnarray}}
\newcommand{\be}{\begin{equation}}
\newcommand{\ee}{\end{equation}}

\newcommand{\bl}{\begin{lem}}
\newcommand{\el}{\end{lem}}
\newcommand{\bp}{\begin{prop}}
\newcommand{\ep}{\end{prop}}
\newcommand{\bt}{\begin{theo}}
\newcommand{\et}{\end{theo}}
\newcommand{\bpr}{\begin{proof}}
\newcommand{\epr}{\end{proof}}

\newcommand{\bc}{\begin{cor}}
\newcommand{\ec}{\end{cor}}

\newcommand{\br}{\begin{rem}}
\newcommand{\er}{\end{rem}}
\newcommand{\bd}{\begin{defi}}
\newcommand{\ed}{\end{defi}}
\newcommand{\bass}{\begin{ass}}
\newcommand{\eass}{\end{ass}}

\newlength{\pecettawidth}
\begin{document}

\begin{frontmatter}


\selectlanguage{english}
\title{Does communication enhance pedestrians transport in the dark?}


\selectlanguage{english}
\author[authorlabel1]{Emilio N. M. Cirillo},
\ead{emilio.cirillo@uniroma1.it}
\author[authorlabel2]{Matteo Colangeli},
\ead{matteo.colangeli@gssi.infn.it}
\author[authorlabel3]{Adrian Muntean}
\ead{a.muntean@tue.nl}

\address[authorlabel1]{Dipartimento di Scienze di Base e Applicate per 
             l'Ingegneria, Sapienza Universit\`a di Roma, 
             via A.\ Scarpa 16, I--00161}
  \address[authorlabel2]{Gran Sasso Science Institute, Via F. Crispi 7, 00167, L'Aquila, Italy}           
\address[authorlabel3]{Department of Mathematics and Computer Science, CASA -- Center for Scientific Computing and Applications, Eindhoven University of Technology, PO Box 513, 5600 MB, Eindhoven, The Netherlands}


\medskip
\begin{abstract}
We study the motion of pedestrians through an obscure tunnel where the lack of visibility hides  the exits.
Using a lattice model, we explore the effects of
communication  on the effective transport properties of the crowd of pedestrians. More precisely, we study the effect of two thresholds
on the structure of the effective nonlinear diffusion coefficient. One threshold models pedestrians's communication efficiency in  the dark, while the other one
describes the tunnel capacity.  Essentially, we note that  if the evacuees show a maximum trust (leading to a fast communication), they tend to quickly find the exit
and hence the collective action tends to prevent the occurrence of disasters.

\vskip 0.5\baselineskip

\selectlanguage{francais}
\noindent{\bf R\'esum\'e}
\vskip 0.5\baselineskip
\noindent
Nous \'etudions la dynamique des movements de foules dans un tunnel  dont la visibilit\'e est tr\`es reduite. Tout en particulier, nous nous int\'er\'essons \'a des tunnels dont les sorties ne sont pas visibles. \'A l'aide de notre mod\`ele -- un automate cellullaire -- nous \'exploitons les \'effets 
de la communication inter--personnelle parmis les pi\'etons sur la structure de la nonlinearit\'e du coefficient de diffusion. Nous mod\'elisons l'\'efficacit\'e de la communication inter--personnelle ainsi que la capacit\'e des sorties \'a l'aide de deux barri\`eres.


\vskip 0.5\baselineskip
\noindent{\small{\it Mots-cl\'es~:} La dynamique des mouvements de foules; des automates cellulaire; un sc\'enario d'\'evacuation; des barri\`eres;}}
\end{abstract}
\end{frontmatter}


\selectlanguage{english}
\section{Introduction}
\label{s:int}
This Note deals with the following evacuation scenario: A possibly large group of pedestrians needs to evacuate a long and obscure tunnel.  The lack of visibility is due to either an electricity breakdown or due to a dense smoke.  The basic modeling assumption is that the pedestrians are equally fit,  do not know each other, and also, are unaware of the precise geometry of the tunnel. We wish to build a lattice model to explore the effects of 
communication  on the effective transport properties of these pedestrians. As modeling tool, we use a particular type of particle system,  known as 
zero range process (abbreviated here ZRP), whose dynamics is 
affected by two thresholds.   One threshold  - called {\em activation threshold} -- models pedestrians's communication efficiency in  the dark, while the other one -- the {\em saturation threshold} -- 
describes the tunnel capacity.  From the modeling point of view, the activation threshold is open to many interpretations. In this Note, we associate the 
size of this threshold not only with the geometric level of the possibility of communication, but also with the willingness and ability of the pedestrians to process the  transmitted information to make a decision towards orientation to a potential exit or choice of speed in the dark. We refer to this as {\em level of trust}. Essentially, a small activation threshold implies in this context a high level of trust.

\begin{figure}
\begin{picture}(400,80)(-70,0)
\setlength{\unitlength}{.026cm}
\thicklines
\put(10,20){\line(1,0){120}}
\put(10,50){\line(1,0){120}}
\thinlines
\put(55,20){\line(0,1){30}}
\put(85,20){\line(0,1){30}}
\put(82.5,32.5){\circle*{5}}
\put(82.5,32.5){\circle{15}}
\put(67.5,40){\circle*{5}}
\put(67.5,40){\circle{15}}
\put(65.5,27.5){\circle*{5}}
\put(65.5,27.5){\circle{15}}
\thicklines
\put(10,70){\line(1,0){120}}
\put(10,100){\line(1,0){120}}
\thinlines
\put(55,70){\line(0,1){30}}
\put(85,70){\line(0,1){30}}
\put(82.5,82.5){\circle*{5}}
\put(82.5,82.5){\circle{15}}
\put(67.5,90){\circle*{5}}
\put(67.5,90){\circle{15}}
\put(65.5,77.5){\circle*{5}}
\put(65.5,77.5){\circle{15}}
\put(72.5,82.5){\circle*{5}}
\put(72.5,82.5){\circle{15}}
\thicklines
\put(210,20){\line(1,0){120}}
\put(210,50){\line(1,0){120}}
\thinlines
\put(255,20){\line(0,1){30}}
\put(285,20){\line(0,1){30}}
\put(258,27.5){\circle*{5}}
\put(258,32.5){\circle*{5}}
\put(258,47.5){\circle*{5}}
\put(282.5,27.5){\circle*{5}}
\put(282.5,37.5){\circle*{5}}
\put(282.5,42.5){\circle*{5}}
\put(282.5,47.5){\circle*{5}}
\thicklines
\put(210,70){\line(1,0){120}}
\put(210,100){\line(1,0){120}}
\thinlines
\put(255,70){\line(0,1){30}}
\put(285,70){\line(0,1){30}}
\put(258,77.5){\circle*{5}}
\put(258,72.5){\circle*{5}}
\put(258,82.5){\circle*{5}}
\put(258,87.5){\circle*{5}}
\put(258,92.5){\circle*{5}}
\put(258,97.5){\circle*{5}}
\put(282.5,77.5){\circle*{5}}
\put(282.5,72.5){\circle*{5}}
\put(282.5,82.5){\circle*{5}}
\put(282.5,87.5){\circle*{5}}
\put(282.5,92.5){\circle*{5}}
\put(282.5,97.5){\circle*{5}}
\put(277,72.5){\circle*{5}}
\put(272.5,87.5){\circle*{5}}
\put(265,97.5){\circle*{5}}
\end{picture}
\caption{Sketch of pedestrians moving through a cell of the obscure tunnel driven by a two-threshold biased dynamics.
}
\label{fig:threshold}
\end{figure}
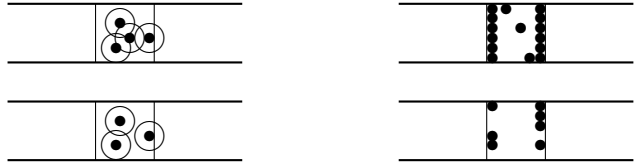

In Fig. \ref{fig:threshold} we sketch the meaning of the two thresholds, the precise mathematical definition is given in Section \ref{s:modello}. Each solid circle represents a pedestrian, whereas the associated
open (bigger) circle represents its communication domain and level of trust. 
On left bottom,  the number of pedestrians in the cell is so small that
their typical distance is larger than the radius of the communication domain.
On left top we see that if the number of pedestrian is large enough information can
propagate throughout the cell.
In other words, we assume that the information can be efficiently 
transmitted among the different pedestrians as soon as 
any single communication domain (the open circles) intersects at least 
one other pedestrian. Essentially, we need here a minimal degree of 
packing of the open circles, which is guaranteed in our scenario 
by the activation threshold. 

On the right bottom part we indicate that provided the number of pedestrians in the cell is smaller than
the number that can be accommodated at boundaries, then it
increases proportionally to
the number of pedestrians in the cell.
If the number of pedestrians in the cell is too high (see right top),
than the number of them exiting the cell per unit time saturates. 

It is worth mentioning that thresholds--biased 
dynamics have been discussed also for other transportation scenarios; compare e.g. \cite{CM01,CM02,CM03} (group formation and cooperation in the dark) and \cite{LH93,Joa06} (collective dynamics of molecular motors).
This Note focusses on communication efficiency and is organized as follows: our transport model is described in Section  \ref{s:modello}, while Section \ref{s:hydlimeq} contains the hydrodynamic limit of the model  as well as numerical illustrations exploring the effects of 
communication  on the effective transport properties of the crowd of pedestrians traveling the obscure tunnel.

\section{The model} 
\label{s:modello}
\par\noindent
We consider a positive integer $L$ and define a 
zero range process (ZRP) \cite{pres,EH} on the finite 
torus (periodic boundary conditions)
$\Lambda:=\{1,\dots,L\}\subset\bb{Z}$. Fix 
$N\in\bb{Z}_+$ and consider the finite \emph{state} or
\emph{configuration space}
$\Omega:=\{0,\dots,N\}^\Lambda$.
Given
$\omega=(\omega_1,\dots,\omega_L)\in\Omega$
the integer $\omega_x$ is called \emph{number of particle}
at site $x\in\Lambda$ in the \emph{state} or \emph{configuration}
$\omega$. We pick $A,S\in\{1,\dots,N\}$ with $S\ge A$, 
the \emph{activation} and \emph{saturation thresholds}, respectively.
We define the \emph{intensity function}
\begin{equation}
\label{soglia}
g(k)
=
\left\{
\begin{array}{ll}
0 &  \textrm{ if } k=0\\
1 & \textrm{ if } 1\leq k\le A\\
k-A+1 & \textrm{ if } A< k\le S\\
S-A+1 & \textrm{ if } k> S\\
\end{array}
\right.
\end{equation}
for each $k\in\bb{Z}_+$. 
The ZRP we consider here 
is the Markov process $\omega_t\in\Omega$, with $t\ge0$, 
such that each 
site $x\in\Lambda$ is updated with intensity $g(\omega_x(t))$
and, once such a site $x$ is chosen, a particle jumps with 
probability $1/2$ to the 
neighboring right site $x+1$ or with probability $1/2$ 
to the neighboring left site $x-1$. 
For more details on this modeling strategy, see \cite{pres}.

The intensity function relates to the 
\textit{hopping rates} 
$r^{(x,x\pm1)}(\omega_x(t))
=
g(\omega_x(t))/2
$
and coincides with the \textit{escape rate} 
$
r^{(x,x-1)}(\omega_x(t))
+
r^{(x,x+1)}(\omega_x(t))
=g(\omega_x(t))
$
at which a particle leaves the site $x$. 
The thresholds intend to control the 
escape rate. Essentially, the activation threshold $A$ 
keeps the escape rate low  for all sites for 
which $\omega_x(t)\le A$, regardless the number of particles on $x$.
The saturation threshold $S$ 
holds the escape rate fixed to a maximum value for all sites for 
which $\omega_x(t)\ge S$, regardless, again, the number of particles on $x$.
In the intermediate case, $A<\omega_x(t)<S$
the escape rate increases proportionally to the actual number of particles on 
$x$, see \eqref{soglia}.  In the limiting case $A=1$ and $S=\infty$, 
the intensity function becomes  
$g(k)=k$, for $k>0$, and thus 
the well-known
\emph{independent particle} model is recovered.
A different limiting situation appears when 
the intensity function is  $1$ for any $k\ge1$ and 
 $0$ for $k=0$. For this we find a ZRP whose configurations can be mapped to a
\emph{simple exclusion}--like model states (cf.\ e.g.\ \cite{EH}). 
Interestingly, we can tune between the two very different dynamics
either 
by keeping $S=\infty$ and varying $A$ or by keeping $A=1$ and 
varying $S$. 

\section{Hydrodynamic limit. Threshold effects}
\label{s:hydlimeq}
\par\noindent
We study
the hydrodynamic limit $N,L\to\infty$. Particularly, we exploit 
the fact that the intensity function is not decreasing and 
use well--established theories to derive 
the limiting non--linear diffusion coefficient and 
the limiting current in presence of the 
two thresholds. The \emph{Gibbs measure} 
with \emph{fugacity} $z\in\mathbb{R}_+$ 
of the ZRP introduced above 
is the product measure 
$
\nu_{z}(\eta_1) 
\nu_{z}(\eta_2) 
\cdots
\nu_{z}(\eta_L) 
$
on $\mathbb{N}^\Lambda$ 
for any 
$\eta=(\eta_1,\dots,\eta_L)\in\mathbb{N}^\Lambda
$
with
$
\nu_{z}(0)=C_{z} 
$
and
$
\nu_{z}(k)=C_{z} z^k/[g(1)\cdots g(k)]
$
for
$k\ge1$,
%
where $C_{z}$ is a normalization factor depending 
on $z$, $A$, and $S$, namely, 
$1/C_z
=
     1+\sum_{k=1}^\infty z^k/[g(1)\cdots g(k)]
.
$
To compute the mean value  
of the intensity function $g$, we use $\nu_{z}(0)$ with $g(0)=0$  to get 
\begin{equation}
\label{mediaI}
\nu_{z}[g(\omega_x)]
=
\sum_{k=1}^\infty \nu_{z}(k)
\,g(k)
=
C_{z} z + C_{z} z \sum_{k=2}^\infty \frac{z^{k-1}}{g(1)\cdots g(k-1)}
=z
\,\,.
\end{equation}
As a function of the activity, 
the expectation does not depend on the particular 
choice of the intensity function. 

For what concerns the hydrodynamic limit, a special role will be 
played by the density 
\be 
\label{dens}
\bar{\rho}(z)
=
\sum_{k=0}^\infty k\,\nu_{z}(k) 
\,.
\ee
Note that 
$\bar\rho(z)$ is an increasing function of the 
fugacity, indeed, 
$
\partial\bar\rho(z)/\partial z
=
[\nu_{z}(\eta_1^2)-(\nu_{z}(\eta_1))^2]/z
>0$. Hence, it is possible to define $\bar z(\rho)$ 
as the inverse function of $\bar\rho(z)$. 
We observe that 
$\bar{\rho}$ is defined for any positive $z$ 
if $A$ is finite and $S=\infty$. 

The evolution of the distribution of the particles 
on the space $\Lambda$ under the ZRP
with thresholds $A$ and $S$ can be described in the diffusive 
hydrodynamic limit via the time evolution of the \textit{density 
function} $\rho(x,t)$ with the space variable $x$ 
varying in the interval $[0,1]$ and for any time $t\ge0$; compare \cite{pres}.
Consequently, the continuous 
space density $\rho(x,t)$ 
is the solution of the partial differential equation 
\be 
\label{diffusion}
\frac{\partial}{\partial t}\rho
=
-\frac{\partial}{\partial x}{J(\varrho)} 
\;\;\textrm{ with } \;\;
J(\varrho)
=
-\frac{1}{2}
D(\rho)
\frac{\partial}{\partial x}\rho
\ee
where the \textit{macroscopic flux} $J(\varrho)$ incorporates the  \textit{effective diffusion coefficient} $D$ given by 
\be 
\label{D}
D(\rho)
=
\frac{\partial }{\partial \rho}
\nu_{\bar{z}(\rho)}\left[g(\omega_1)\right] 
\,.
\ee
The coefficient $D$ is computed in terms 
of the mean of the intensity function evaluated against the 
single site Gibbs measure with fugacity corresponding to the 
local value of the density. 
Consequently, 
$D$ 
depends of the value of the thresholds.

Let us discuss the effects induced by 
the two thresholds $A$ and $S$ on the diffusion coefficient.
We first recall some known results which are valid in the 
two limiting situations
$A=1$ and $S=\infty$ ({\em independent particle model})
and 
$A=S$ ({\em simple exclusion--like model}).
In the first case, 
one has 
$C_z=\exp\{-z\}$. Hence, by \eqref{dens}, it holds
$\bar{\rho}(z) = z$.
Recalling \eqref{mediaI} and 
the definition of $\bar{z}$, 
we have  
$\nu_{\bar{z}(\rho)}\left[g(\omega_1)\right] 
=
\nu_{\rho}\left[g(\omega_1)\right] 
=
\rho
$.
By using \eqref{D},
the diffusion coefficient reads
$D(\rho)
=1$.
On the other hand, in the latter case, 
one has $g(k)=1$ for any $k\ge1$ and $g(0)=0$.
Hence, $C_z=1-z$, and it holds 
$\bar{\rho}(z) = z/(1-z)$.
Here, one finds the law
$D(\rho)=1/(1+\rho)^2$, cf. \cite{Ferrari}.
Now, 
we illustrate the general strategy to compute the diffusion coefficient 
$D$ for arbitrary values of the thresholds $A$ and $S$.
To do so, we first compute 
$\bar{\rho}(z)$. 
The precise expression of the diffusion coefficient can be then obtained 
using the general recipe in equation \eqref{D} and recalling \eqref{mediaI}.
Indeed, 
\begin{equation}
\label{Dgen}
D(\rho)
=
\frac{\partial}{\partial\rho}
\nu_{\bar{z}(\rho)}\left[g(\omega_1)\right] 
=
\frac{\partial}{\partial\rho}
\bar{z}(\rho)
=
\Big(
\frac{\partial}{\partial z}
\bar{\rho}(z)
\Big)^{-1}
\Big|_{z=\bar{z}(\rho)}
\end{equation}
The explicit expression of 
$\partial\bar\rho(z)/\partial z$ appearing in 
\eqref{Dgen} is 
lengthy, hence we omit it. 

\begin{figure}
\centering
\includegraphics[width=5.0cm]{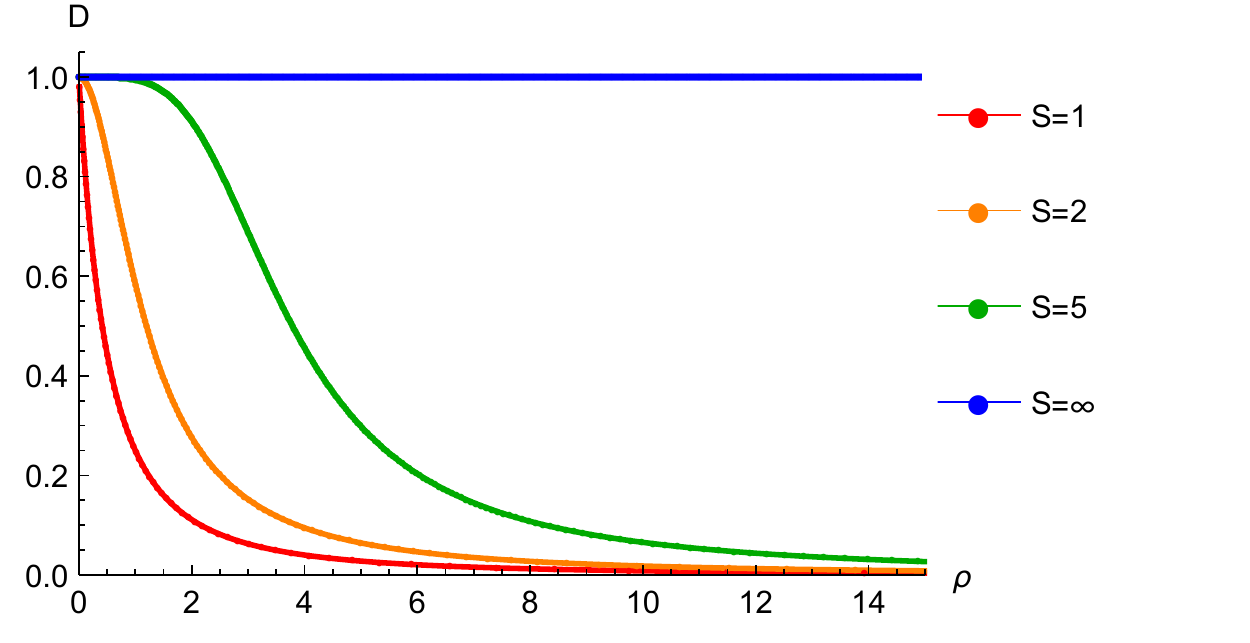}
\hspace{1mm}
\includegraphics[width=5.0cm]{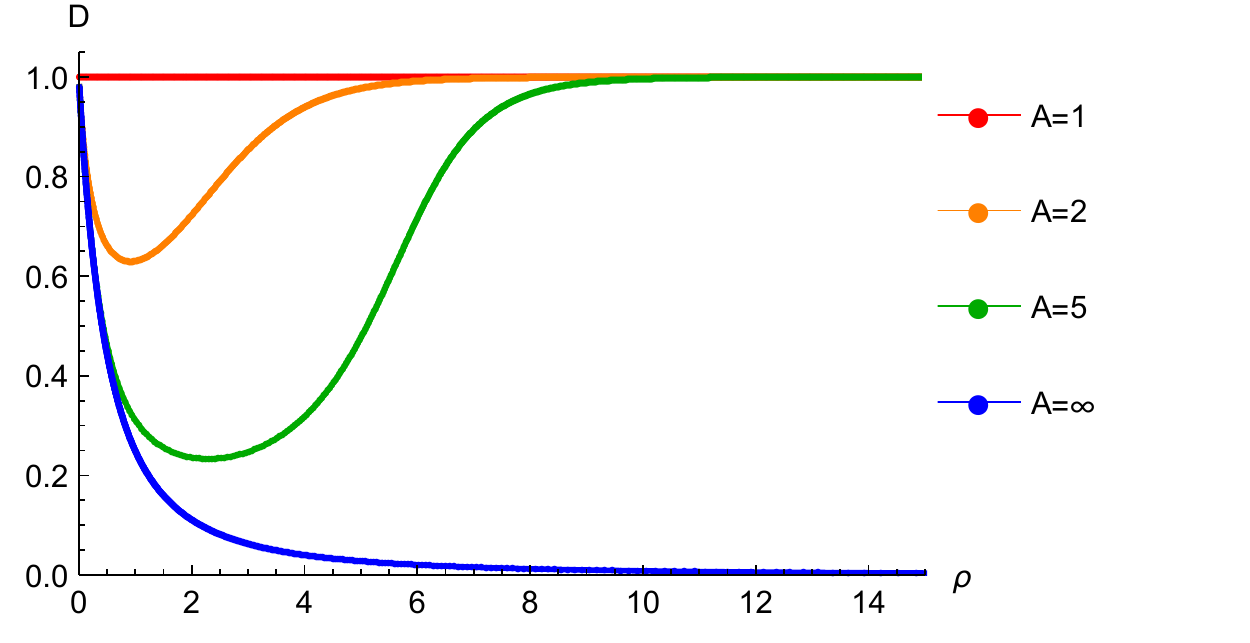}
\vspace{1mm}
\includegraphics[width=5.0cm]{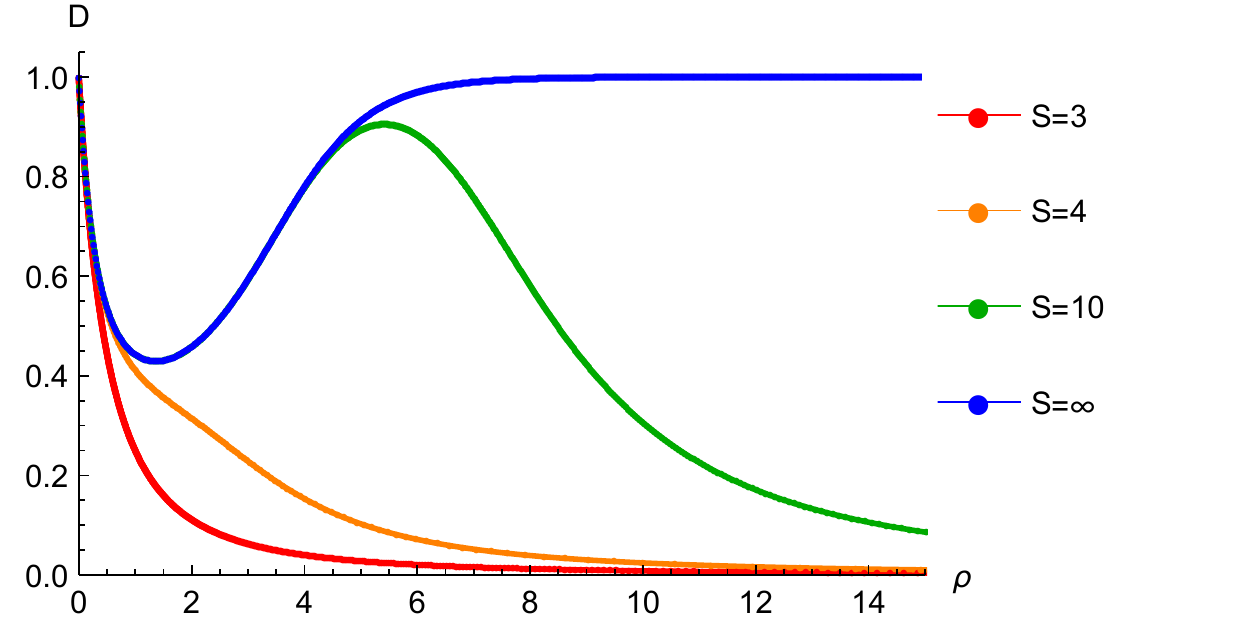}
\hspace{1mm}
\includegraphics[width=5.0cm]{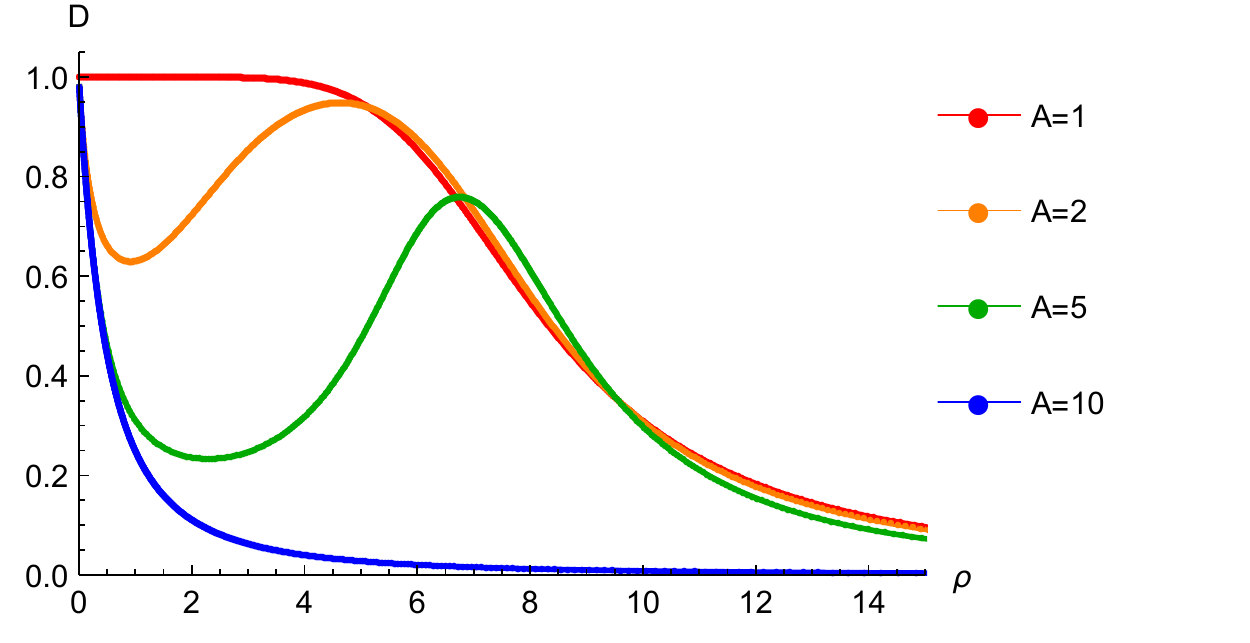}
\caption{\textit{Left panel, top row}: Diffusion coefficient 
$D(\rho)$ vs. $\rho$ for $A=1$ and for different values of the saturation threshold, 
i.e., $S=1,2,5,\infty$. 
\textit{Right panel, top row}: Diffusion coefficient 
$D(\rho)$ vs. $\rho$ for $S=\infty$ and for 
different values of the activation threshold, 
i.e., $A=1,2,5,\infty$. 
\textit{Left panel, bottom row}: Diffusion coefficient 
$D(\rho)$ vs. $\rho$ for $A=3$ and for $S=3,4,10,\infty$. \textit{Right panel, bottom row}: Diffusion coefficient 
$D(\rho)$ vs. $\rho$ for $S=10$ and for $A=1,2,5,10$.}
\label{fig:diff}
\end{figure}

Fig.~\ref{fig:diff} shows the behavior of the diffusion 
coefficient as a function of the local density, and parametrized by the 
values 
of the thresholds. The upper left panel 
refers to the case $A=1$ 
and for different 
values of $S$: 
the simple exclusion--like model is recovered for $S=1$, 
while the independent particle model appears for $S=\infty$. Similarly, the upper right panel illustrates the case with $S=\infty$ 
and for different values of $A$: here the independent particle model corresponds to $A=1$ and the simple
exclusion--like model is recovered for $A=\infty$. 
In both the upper panels of the figure 
we see that for the independent particle 
case, the diffusion coefficient is 
constant with respect to the local density. 

Furthermore, in Fig.~\ref{fig:diff}, we remark 
 the loss of monotonicity of the function $D(\rho)$ for values of $\rho$ exceeding 
some critical value (depending on the thresholds $A$ and $S$). 
The behavior of $D(\rho)$ displayed in the lower left panel 
refers to the case $A=3$.
Considering particularly the green curve  corresponding to $S=10$, one sees 
the onset of a double loss of monotonicity of the function $D(\rho)$: 
for small values of the density, $D$ stays close to the simple exclusion--like behavior and 
decreases with $\rho$. 
After one first critical value of the density, it starts rising up, until it 
drops down again, when $\rho$ exceeds an upper critical value.
This indicates the presence of a double threshold for the intensity function given  by \eqref{soglia}.

In the lower right panel of Fig.~\ref{fig:diff},
the diffusion coefficient is plotted for different values of $A$
for the case $S=10$.
The red and the blue curves 
refer to the extreme independent--particle and simple--exclusion--like cases.
When the activation threshold is varied two effects are prominent: 
the non--monotonicity of the diffusion coefficient with respect to 
density shows up and, at fixed local density, the diffusion coefficient 
decreases when $A$ is increased. Interestingly, note that 
for  local densities close to $8$ this appears not to be true: 
increasing the activation threshold induces an increase in the diffusion 
coefficient. For this very particular regime 
the dynamics accelerates due to the increase in the 
activation threshold. In other words, it seems that close to the local density $8$ (which, depending on the population type, is close to the maximum pedestrians density  before asphyxiation starts off), higher mistrust speeds up the dynamics.  

Excepting the just mentioned non--intuitive regime, 
the effect of the two thresholds on the diffusion coefficient can be 
summarized as follows: {\em the smaller the 
activation threshold $A$ and/or the higher the saturation threshold $S$, 
the higher is the diffusion coefficient, and therefore, the quicker the dynamics.} From the evacuation viewpoint, 
``a small activation threshold increases the diffusion coefficient" means that 
\begin{quote}
\framebox{
higher trust among pedestrians improves
communication in the dark 
}
\end{quote}
and therefore the exits can be found easier. The size of the saturation threshold simply decides on the exit capacity.  Consequently, a higher saturation threshold leads to an improved capacity of the exists (e.g. larger doors, or more exits \cite{Ronchi}) and therefore the evacuation rate is correspondingly higher. 


\paragraph*{Acknowledgements.} We thank E. Presutti (GSSI
L'Aquila, Italy), A. De Masi  (L'Aquila, Italy), 
and C. Landim  (IMPA, Brazil) for useful discussions. 
ENMC thanks ICMS (TU/e, Eindhoven, The Netherlands) for the very
kind hospitality and for financial support.



\end{document}